\documentclass[journal]{IEEEtran}
\usepackage{setspace}
\usepackage[section]{placeins}
\usepackage{graphicx,subfigure}
\usepackage{amsmath}
\usepackage{amsfonts}
\usepackage{caption2}
\usepackage{amssymb}
\usepackage{varwidth}
\usepackage{multirow}
\usepackage{booktabs}
\usepackage[usenames, dvipsnames]{color}
\usepackage[linesnumbered,ruled,vlined]{algorithm2e}
\usepackage{array}
\usepackage{rotating}
\usepackage{epsfig,amsfonts,multirow}
\usepackage{bm}
\usepackage{cite}

\usepackage{pifont}
\newcommand{\cmark}{\ding{51}}%
\newcommand{\xmark}{\ding{55}}%
\begin{document}
\title{Hyperspectral Image Classification with Attention Aided CNNs}
\author{Renlong~Hang,~\IEEEmembership{Member,~IEEE}, Zhu Li,~\IEEEmembership{Senior~Member,~IEEE}, Qingshan~Liu,~\IEEEmembership{Senior~Member,~IEEE}, \\ Pedram Ghamisi,~\IEEEmembership{Senior~Member,~IEEE},
and Shuvra S. Bhattacharyya,~\IEEEmembership{Fellow,~IEEE}

\thanks{R. Hang is with the School of Automation, Nanjing University of Information Science and Technology, Nanjing 210044, China, and also with the Department of Computer Science and Electrical Engineering, University of Missouri-Kansas City, Missouri 64110, USA (e-mail: renlong\_hang$@$163.com).

Z. Li is with the Department of Computer Science and Electrical Engineering, University of Missouri-Kansas City, Missouri 64110, USA (e-mail: lizhu@umkc.edu).

Q. Liu is with the Jiangsu Key Laboratory of Big Data Analysis Technology, the School of Automation, Nanjing University of Information Science and Technology, Nanjing 210044, China (e-mail: qsliu$@$nuist.edu.cn).

P. Ghamisi is with the Helmholtz-Zentrum Dresden-Rossendorf (HZDR), Helmholtz Institute Freiberg for Resource Technology (HIF), Exploration, D-09599 Freiberg, Germany (e-mail: p.ghamisi@gmail.com).

S. S. Bhattacharyya is with the Department of Electrical and Computer Engineering, University of Maryland, College Park, MD 20742, USA (e-mail: ssb@umd.edu).

}}

\maketitle

\begin{abstract}
Convolutional neural networks (CNNs) have been widely used for hyperspectral image classification. As a common process, small cubes are firstly cropped from the hyperspectral image and then fed into CNNs to extract spectral and spatial features. It is well known that different spectral bands and spatial positions in the cubes have different discriminative abilities. If fully explored, this prior information will help improve the learning capacity of CNNs. Along this direction, we propose an attention aided CNN model for spectral-spatial classification of hyperspectral images. Specifically, a spectral attention sub-network and a spatial attention sub-network are proposed for spectral and spatial classification, respectively. Both of them are based on the traditional CNN model, and incorporate attention modules to aid networks focus on more discriminative channels or positions. In the final classification phase, the spectral classification result and the spatial classification result are combined together via an adaptively weighted summation method. To evaluate the effectiveness of the proposed model, we conduct experiments on three standard hyperspectral datasets. The experimental results show that the proposed model can achieve superior performance compared to several state-of-the-art CNN-related models.

\end{abstract}
\IEEEpeerreviewmaketitle

\section{Introduction}
Similar to the semantic segmentation task of natural images, hyperspectral image classification aims at assigning one of pre-defined categories to each pixel. An important issue for this task is how to represent each pixel effectively. Hyperspectral sensors are capable of capturing the spectral signature of each material along different spectral bands \cite{xu2019nonlocal, hong2018augmented}.
This rich spectral information is able to be used as a feature representation for each pixel in hyperspectral images \cite{ghamisi2017advanced}. Due to the existence of the spectral variability between the same class materials and the spectral similarity between the different class materials, the individual use of spectral features will easily cause misclassified pixels, which is recognized as the ``salt and pepper'' noise in the classification maps. In addition to spectral information, hyperspectral images also contain spatial information. Taking advantage of spectral and spatial features jointly can dramatically alleviate the aforementioned misclassification issue, thus becoming a hot research topic in the field of hyperspectral image classification \cite{he2018recent, ghamisi2018}.

Most of the traditional feature extraction techniques depend on pre-designed criterions by human experts \cite{kang2020learning, hong2019cospace}. For example, Hong \textit{et al.} \cite{hong2019cospace} for the first time took the remote sensing image classification as a cross-modality learning problem, and proposed a milestone common subspace learning algorithm. However, it is difficult to thoroughly explore the intrinsic properties of data.
In recent years, deep learning has demonstrated its overwhelming superiority in numerous computer vision fields \cite{lecun2015}. In contrast with the traditional feature extraction techniques, it combines the task of feature extraction and that of image classification into a unified framework, and lets the data itself drive the optimization of this end-to-end model, thus achieving more robust and discriminative features. Due to its powerful feature learning ability, deep learning has been naturally employed to the classification task of hyperspectral images \cite{zhu2017, li2019deep}. Typical deep learning models include autoencoders \cite{chen2014deep, tao2015unsupervised, ma2016spectral}, recurrent neural networks (RNNs) \cite{liu2017bidirectional, zhou2019hyperspectral, hang2019cascaded}, and convolutional neural networks (CNNs) \cite{chen2016deep, li2017hyperspectral, hang2020classification}. The inputs of autoencoders and RNNs are vectors. Therefore, they can be easily adopted for spectral feature extraction, but will lose some useful information when applied to extract spatial features. In comparison with them, CNNs are able to deal with both spectral feature and spatial feature flexibly, thus becoming the most popular deep learning model for hyperspectral image classification.

According to the input information of networks, existing CNN models are able to be grouped into two classes: spectral CNNs and spectral-spatial CNNs. Spectral CNNs aim at extracting spectral features for each pixel in hyperspectral images. For example, in \cite{hu2015deep}, Hu \textit{et al.} designed a 1-D CNN model for extracting features from the spectral information of each pixel. It mainly consists of one convolutional layer and two fully-connected layers. Since there often exist small numbers of training pixels, the proposed CNN model is not very deep, which limits the feature representation ability of 1-D CNNs. In order to address this issue, a novel pixel-pair method was proposed in \cite{li2017hyperspectral}. By regarding the pixel learning problem as the pixel-pair counterpart, the number of training pixels is significantly increased. Thus, a deeper 1-D CNN model with ten convolutional layers was successfully trained, improving the spectral classification results compared to the shallow CNN model \cite{hu2015deep}. In \cite{wu2017convolutional} and \cite{wu2018semi}, Wu and Prasad proposed to combine 1-D CNN and RNN together. Specifically, they fed the spectral features learned by a 1-D CNN into a RNN to further fuse and enhance the discriminative ability of the extracted features.

Different from spectral CNNs, the purpose of spectral-spatial CNNs is extracting spectral and spatial features simultaneously from hyperspectral images. An intuitive method is adopting 3-D convolutional kernels to construct the CNN model \cite{chen2016deep, li2017spectral, paoletti2018deep}, so the rich spectral and spatial information can be integrated together in each convolutional layer. However, these 3-D convolutional operators often cost much time or need large numbers of parameters. To alleviate this problem, a lot of works attempted designing two-branch networks. One of them focuses on spectral feature extraction, and the other one aims at spatial feature extraction. These results are then combined together using different kinds of fusion strategies. For example, in \cite{yang2017learning} and \cite{xu2018multisource}, a parallel two-branch framework was proposed, where a 1-D CNN and a 2-D CNN were designed to extract spectral and spatial features, respectively. For the 2-D CNN model, its inputs were constructed by extracting a few principal components \cite{zhao2016spectral}, thus the computational consuming was significantly reduced as compared to 3-D CNNs. In \cite{zhong2018spectral}, a serial two-branch framework was designed. It firstly applied several $1\times1$ convolutions to extract spectral features and then fed them into several 2-D convolutions to extract spatial features.

Similar to traditional feature extraction techniques, spectral-spatial CNNs often acquire superior classification performance than spectral CNNs because of the joint exploitation of spectral and spatial information. Therefore, we focus on spectral-spatial CNNs in this paper. In general, small cubes are firstly cropped from the hyperspectral image and then fed into spectral-spatial CNNs to extract features. However, it is well known that different spectral bands and spatial positions in the cubes have different discriminative abilities. If fully explored, this prior information will help improve the learning capacity of CNNs. Recently, attention mechanism has been popularly employed to language modelling \cite{mnih2014recurrent, xu2015show, yang2016stacked} and computer vision tasks \cite{wang2017residual, hu2018squeeze, woo2018cbam}. Its success mainly depends on the reasonable assumption that human vision tends to only focus on selective parts of the whole visual space when and where needed \cite{chen2017sca}. Very recently, similar works have been explored in the field of hyperspectral image processing \cite{fang2019hyperspectral, haut2019visual, mei2019spectral}.
Inspired from them, we propose an attention aided spectral-spatial CNN model for hyperspectral image classification. Our goal is to enhance the representation capacity of CNNs by using attention mechanism, making CNNs focus on more discriminative spectral bands and spatial positions while suppress unnecessary ones.

During the last few years, numerous attention models have been developed. In \cite{wang2018non}, Wang \textit{et al.} designed a spatial attention module. The response value at each position was derived according to the weighted summation of the features at all positions. They used several $1\times1$ convolutions to achieve this goal. In \cite{hu2018squeeze}, Hu \textit{et al.} proposed a channel attention module via two fully-connected layers to adaptively recalibrate channel-wise feature responses. For hyperspectral image classification, there only exists a limited number of training samples, so lightweight attention modules are preferred. In \cite{woo2018cbam}, a convolutional layer was employed to construct a spatial attention module. Motivated by it, we also use small convolutional layers to design our spectral and spatial attention modules. Specifically, our spatial attention module is mainly comprised by one $1\times1$ convolution and two small convolutions. The goal of the $1\times1$ convolution is to reduce the channel numbers in 3-D feature maps to 1. Similar to the spatial attention module, our spectral attention module is mainly comprised of an average pooling layer and two small convolutional layers. The average pooling layer aims at reducing the spatial size in 3-D feature maps to $1\times1$. More importantly, in both spectral and spatial attention modules, we use an output layer to aid them learn more discriminative features. Based on these two kinds of attention modules, we are able to construct two attention sub-networks. One of them incorporates spectral attention modules into a 2-D CNN for extracting spectral features and classification, while the other one incorporates spatial attention modules into another 2-D CNN for extracting spatial features and classification. Our major contributions can be summarized as follows.
\begin{enumerate}
  \item We propose a two-branch spectral-spatial attention network for hyperspectral image classification. Compared to the existing CNNs, our model incorporates attention modules to each convolutional layer, making CNNs focus on more discriminative channels and spatial positions, while suppress unnecessary ones. In the classification phase, two-branch results are fused together via an adaptively weighted summation method.
  \item Considering the limited numbers of training samples, we propose a lightweight spectral attention module via two convolutional operators. Before convolutional layers, we use global average pooling to reduce the effects of spatial information. More importantly, we add an output layer in the module to aid its learning process.
  \item Similar to the spectral attention module, we also use two convolutional layers to construct the spatial attention module. Instead of using pooling operators, we adopt one $1\times1$ convolutional layer for reducing the number of channels to 1. Also, an output layer is added to guide the learning process of the spatial module.
\end{enumerate}

The following sections are organized as follows. Section II presents the proposed model in detail, including the structure of CNNs, the attention modules, and the network training method. Section III describes the experimental data and results. Finally, the conclusions of this paper are summarized in Section IV.

\section{Methodology}
\subsection{Framework of the Proposed Model}
\begin{figure*}
  \centering
  \includegraphics[scale=0.65]{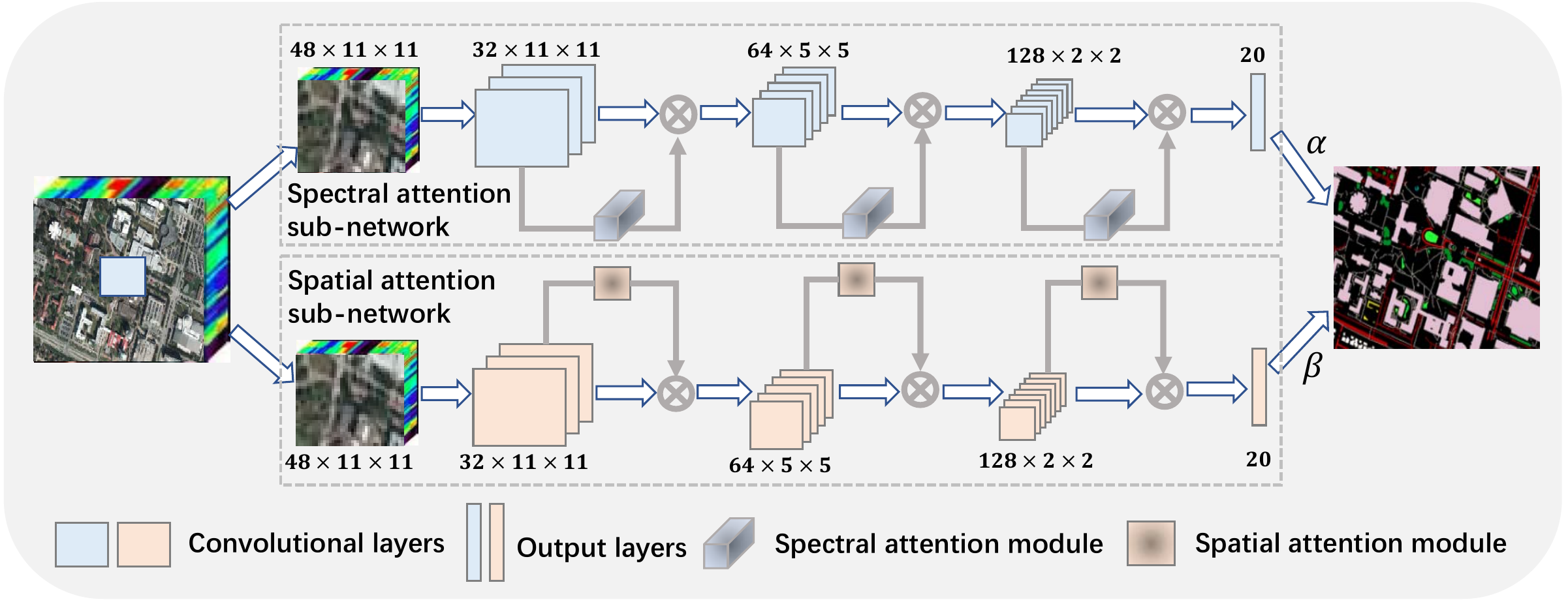}\\
  \caption{Flowchart of the proposed model. Note that the numbers represent the size of each layer for the Houston 2018 data.}\label{Flowchart}
\end{figure*}

Fig.$~$\ref{Flowchart} shows the flowchart of the proposed model. It mainly consists of two branches: the spectral attention sub-network and the spatial attention sub-network. Different from the widely used CNNs, the spectral and spatial attention sub-networks incorporate attention modules to refine the feature maps in each convolutional layer, thus enhancing the learning ability of CNNs. Specifically, for a given pixel, a small cube centered at it is firstly extracted. Then, the cube is fed into the spectral attention sub-network and the spatial attention sub-network simultaneously to obtain two classification results. Finally, a weighted summation method is employed to combine these two results together. Assume that there are $K$ classes to discriminate, $\mathbf{O}_{spe}\in\Re^{K}$ and $\mathbf{O}_{spa}\in\Re^{K}$ represent the output results of the spectral attention sub-network and the spatial attention sub-network, respectively. The final output $\mathbf{O}\in\Re^{K}$ of the proposed model is:
\begin{equation}\label{output}
\begin{aligned}
& \mathbf{O} = \alpha\times\mathbf{O}_{spe} + \beta\times\mathbf{O}_{spa}
& s.t. \;\,\alpha + \beta = 1
\end{aligned}
\end{equation}
where $\alpha$ and $\beta$ are the weighting parameters. They can be adaptively learned during the optimization process of the whole network. The $i$th element in $\mathbf{O}$ denotes the probability that the given pixel is classified as the $i$th category.

\subsection{The Structure of CNNs}
As demonstrated in Fig.$~$\ref{Flowchart}, the spectral attention sub-network and the spatial attention sub-network consist of CNNs and attention modules. In this subsection, we will present the basic structure of the adopted CNNs. Then, in the next subsection, we will describe the attention modules in detail.

For hyperspectral images, there only exists a limited number of training samples, so it is difficult to train very deep CNNs. The same as \cite{chen2016deep}, we also apply three convolutional layers to construct the spectral and spatial sub-networks. Each convolutional layer is sequentially followed by a batch normalization layer to regularize and accelerate the training process, and a rectified linear unit (ReLU) to learn a nonlinear representation. Before the second and the third convolutional layer, a max-pooling layer is adopted for reducing the data variance and the computation complexity. The kernel size of each convolutional layer is $3\times3$, and the channel numbers from the first to the third convolution layer are 32, 64, and 128, respectively.

For the $l$th convolutional layer, the $i$th feature map can be represented as:
\begin{equation}\label{Conv}
  \mathbf{F}_{i}^{l} = f(\sum_{j}\mathbf{F}_{j}^{l-1}\ast\mathbf{w}_{i,j}^{l} + b_{i}^{l})
\end{equation}
where $l\in\{1,2,3\}$, $\mathbf{F}_{j}^{l-1}$ is the $j$th feature map at the $(l-1)$th layer, $\mathbf{F}_{j}^{0}$ is the $j$th spectral band of the original input cube, $\mathbf{w}_{i,j}^{l}\in\Re^{3\times3}$ is the convolutional kernel, `$\ast$' is the convolutional operator, $b_{i}^{l}$ is the bias, and $f$ is the ReLU activation function. Note that the spatial size of $\mathbf{F}^{l}$ is the same as that of $\mathbf{F}^{l-1}$ via a padding operator. In Equation$~$(\ref{Conv}), the convolutional operator and the summation operator can learn the spatial features and aggregate the spectral features, respectively.

\subsection{Attention Modules}

It is well known that the spectral responses at different bands may vary largely for the same object, which means the discriminative abilities of different bands are diverse. In addition, different positions of the extracted cube also have different semantic information. For example, the object edges are generally more discriminative than the other positions. If such prior information can be fully explored, the learning ability of the spectral and spatial sub-networks will be improved. In this paper, we design two classes of attention modules to achieve this goal. They are the spectral attention module and the spatial attention module. The spectral attention module is adopted to make the spectral sub-network focus on more discriminative channels while suppress unnecessary ones. Similarly, the spatial attention module can make the spatial sub-network pay more attention to the semantic positions.

\begin{figure}
  \centering
  \includegraphics[scale=0.6]{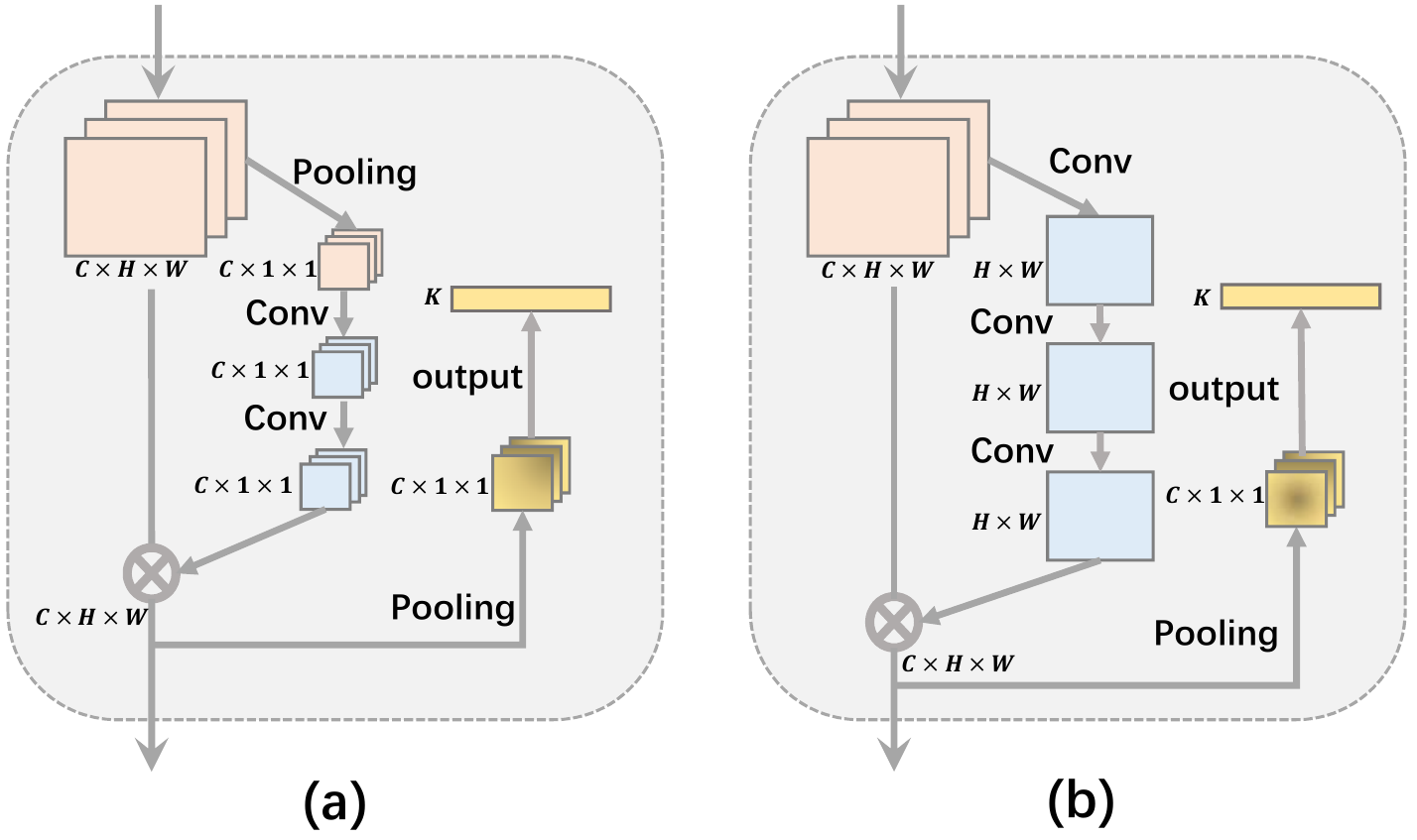}\\
  \caption{Attention modules. (a) Spectral attention module. (b) Spatial attention module. }\label{Attention}
\end{figure}

\textbf{Spectral Attention.} As shown in Fig.$~$\ref{Attention}(a), the spectral attention module is constructed by exploiting the inter-channel relationships of feature maps. Given an intermediate feature map $\mathbf{F}^{l}\in\Re^{C\times H\times W}$, where $l\in\{1, 2, 3\}$, $C$, $H$, and $W$ represent channel numbers, the height, and the width of $\mathbf{F}^{l}$, respectively, a global average-pooling layer is firstly applied to squeeze the spatial dimension of it. Then, two 1-D convolutional layers are employed to generate a spectral attention map $A_{spe}(\mathbf{F}^{l})\in\Re^{C\times 1\times 1}$, which can be formulated as follows:
\begin{equation}
 A_{spe}(\mathbf{F}^{l}) = \sigma(\mathbf{W}_{2}\ast f(\mathbf{W}_{1}\ast \mathbf{F}^{l}_{avg}))
\end{equation}
where $\sigma$ denotes a sigmoid function, $f$ is the ReLU function, $\mathbf{F}^{l}_{avg}\in\Re^{C\times 1\times 1}$ represents the feature map obtained by the global average-pooling operator, $\mathbf{W}_{1}\in\Re^{k\times 1\times 1}$ and $\mathbf{W}_{2}\in\Re^{k\times 1\times 1}$ are the first and the second convolution kernels, respectively. Note that padding operators are employed in each convolutional layer for making the output size equal to $C$. Since $C$ increases when $l$ changes from 1 to 3, larger $k$ values are used as $l$ increases. Specifically, $k$ is set to 3, 5 and 7 when $l$ equals to 1, 2, and 3, respectively.

After acquiring $A_{spe}(\mathbf{F}^{l})$, we apply it to refine the original feature map $\mathbf{F}^{l}$ as follows:
\begin{equation}
(\mathbf{F}^{l})^{\prime} = \mathbf{F}^{l}\otimes A_{spe}(\mathbf{F}^{l})
\end{equation}
where `$\otimes$' represents element-wise multiplication. During multiplication, the values in $A_{spe}(\mathbf{F}^{l})$ are expanded (copied) along the spatial dimension. Finally, $(\mathbf{F}^{l})^{\prime}$ is adopted as an input for the $(l+1)$th convolutional layer and an output branch (yellow colors in Fig.$~$\ref{Attention}). The output branch is comprised of a global max-pooling layer and a fully-connected layer. This branch mainly has two purposes: the first one is to provide supervised information for the spectral attention module, ensuring the discriminative ability of the refined feature map; the other one is to incorporate a regularization term to the loss function (discussed in subsection II-D), alleviating the overfitting problem during network training.

\textbf{Spatial Attention.} Similar to the spectral attention module, the spatial attention module is built by taking advantage of the inter-spatial relationships of feature maps. As demonstrated in Fig.$~$\ref{Attention}(b), an $1\times1$ convolutional layer is firstly used to aggregate the information along the channel direction of $\mathbf{F}^{l}$, generating a 2-D feature map $\mathbf{M}^{l}\in\Re^{H\times W}$. Then, two 2-D convolutional layers are applied to derive a spatial attention map $A_{spa}(\mathbf{F}^{l})\in\Re^{H\times W}$, which can be formulated as:
\begin{equation}
 A_{spa}(\mathbf{F}^{l}) = \sigma(\mathbf{Q}_{1}\ast f(\mathbf{Q}_{2}\ast \mathbf{M}^{l}))
\end{equation}
where $\mathbf{Q}_{1}\in\Re^{k\times k}$ and $\mathbf{Q}_{2}\in\Re^{k\times k}$ have the same kernel size. Also, padding operators are employed at each convolutional layer to avoid the change of spatial sizes. Due to the decreasing of spatial sizes as $l$ increases, $k$ is set to 7, 5 and 3 when $l$ equals to 1, 2, and 3, respectively. After that, $A_{spa}(\mathbf{F}^{l})$ can be used to recalibrate $\mathbf{F}^{l}$ as:
\begin{equation}
(\mathbf{F}^{l})^{\prime} = \mathbf{F}^{l}\otimes A_{spa}(\mathbf{F}^{l})
\end{equation}
During multiplication, the values in $A_{spa}(\mathbf{F}^{l})$ are expanded (copied) along the channel dimension. Finally, $(\mathbf{F}^{l})^{\prime}$ is fed into the $(l+1)$th convolutional layer and an output branch. Here, we use an adaptive max-pooling layer in the output branch, which means that the output size is fixed for any size inputs. Specifically, the output size is fixed to $4\times4$, $2\times2$, and $1\times1$ when $l$ equals to 1, 2, and 3, respectively.

\subsection{Network Training}
To train the proposed model effectively, we adopt a two-step strategy. The first step is pre-training the two sub-networks independently, while the second step is adding the weighted summation layer and fine-tuning the whole network. Assume the output result of the $i$th attention module for the $j$th training sample is $\mathbf{O}^{(j)}_{i}\in\Re^{K}, i\in\{1,2,3\}$, and its ground-truth label is $\mathbf{y}^{(j)}$, so the loss value can be formulated as:
\begin{equation}\label{Loss}
  L = \sum_{j=1}^{N}\sum_{i=1}^{3} \gamma_{i}\times L(\mathbf{O}^{(j)}_{i}, \mathbf{y}^{(j)})
\end{equation}
where $N$ denotes the total number of training pixels, and $L(\cdot,\cdot)$ represents the loss function. Without loss of generality, the cross-entropy loss function is chosen. Since the deeper convolutional layers are expected to capture more discriminative features, we set larger weights for them. Thus, $\gamma_{1}$, $\gamma_{2}$, and $\gamma_{3}$ are empirically set to 0.01, 0.1, and 1, respectively. During the pre-training process, we choose the gradient descent algorithm to optimize $L$. After that, we apply Equation$~$(\ref{output}), where $\mathbf{O}_{spe}$ and $\mathbf{O}_{spa}$ are replaced by their respective $\mathbf{O}^{(j)}_{3}$, to re-calculate the output value $\mathbf{O}^{(j)}$. Based on $\mathbf{O}^{(j)}$, we can update the loss value $L$ as follows:
\begin{equation}
\widetilde{L} = \sum_{j=1}^{N} L(\mathbf{O}^{(j)}, \mathbf{y}^{(j)})
\end{equation}
Again, the gradient descent algorithm is used to optimize $\widetilde{L}$ during the fine-tuning process.

\section{Experiments}
\subsection{Data Description and Experimental Setup}
\begin{figure}
  \centering
  \includegraphics[scale = 0.5]{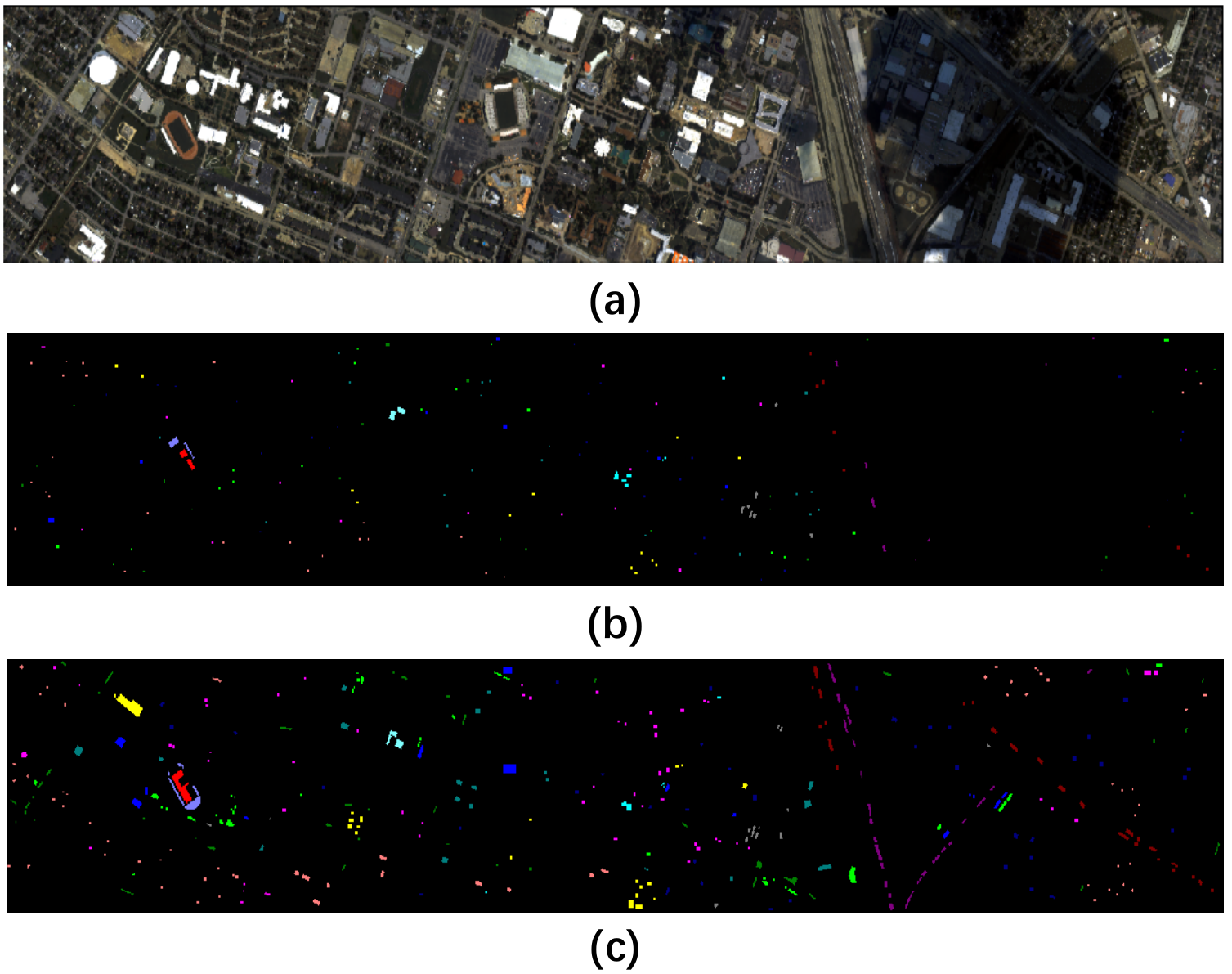}\\
  \caption{Houston 2013 data visualization. (a) False-color image. (b) Training data visualization. (c) Test data visualization.}\label{RGBHS}
\end{figure}

\begin{figure}
  \centering
  \includegraphics[scale = 0.5]{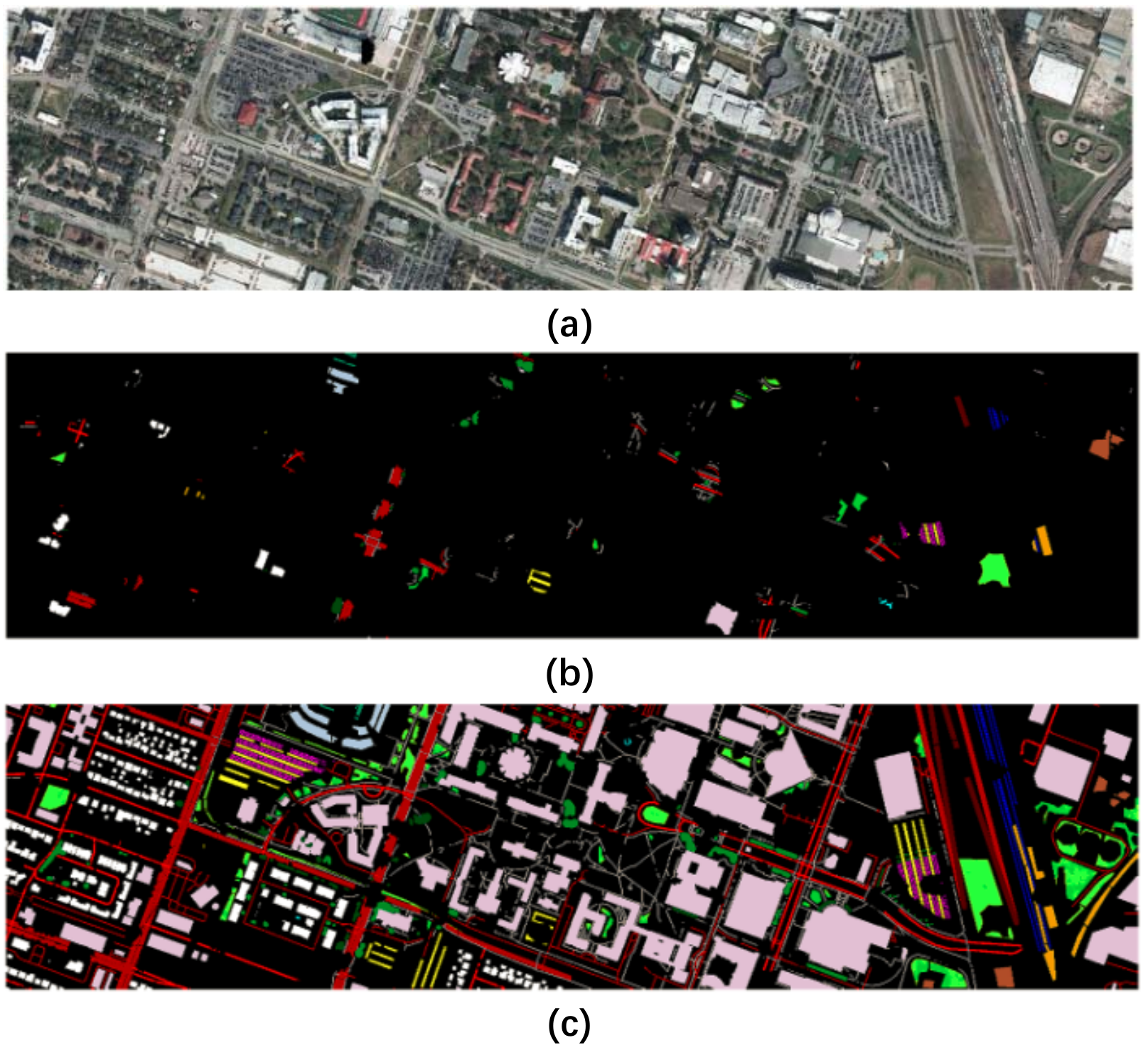}\\
  \caption{Houston 2018 data visualization. (a) False-color image. (b) Training data visualization. (c) Test data visualization.}\label{RGBHU2018}
\end{figure}

\begin{figure}
  \centering
  \includegraphics[scale = 0.45]{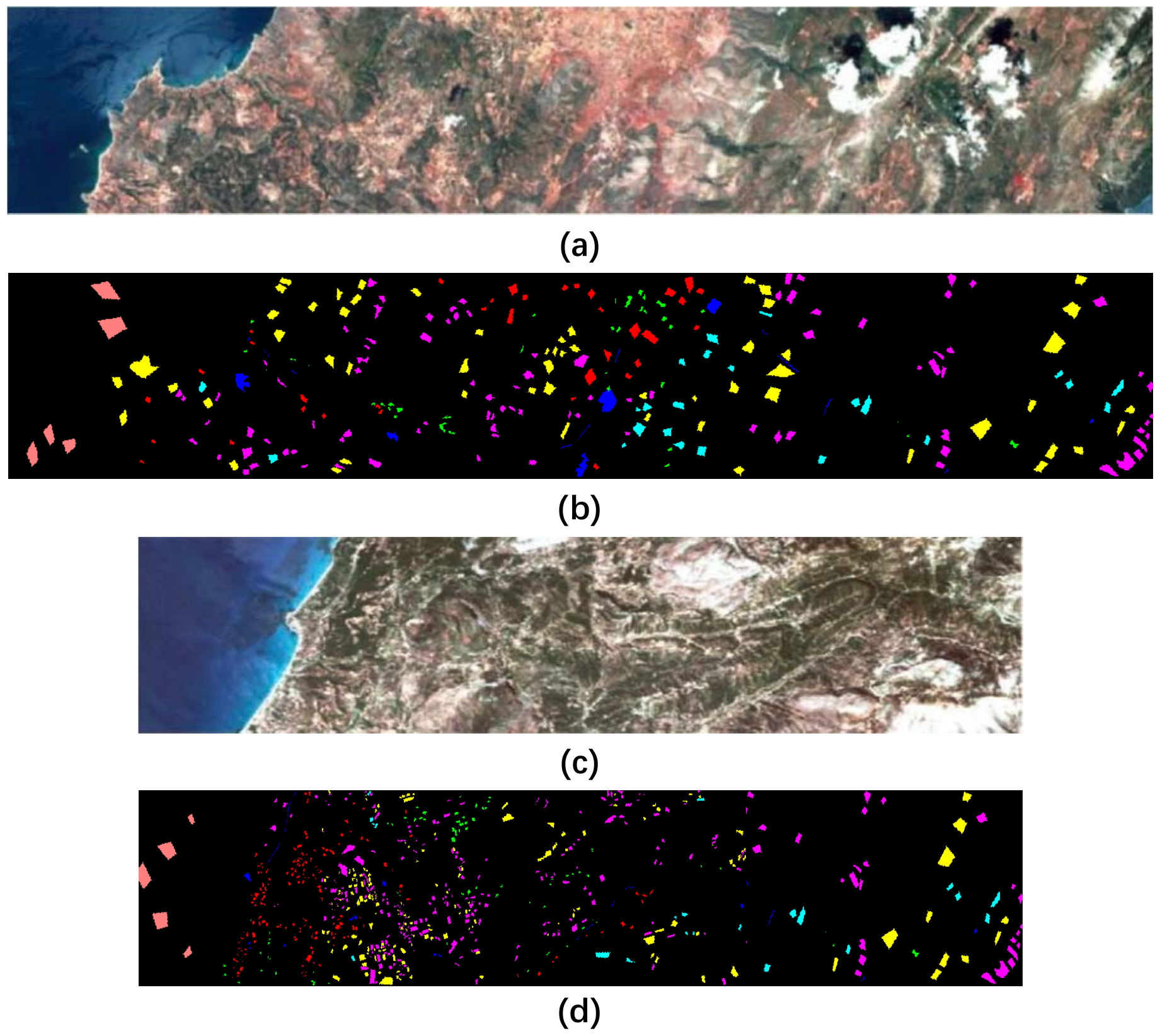}\\
  \caption{HyRANK data visualization. (a) and (b) are the false-color image and the available pixel map on the Dioni data. (c) and (d) are the false-color image and the available pixel map on the Loukia data.}\label{RGBHyRANK}
\end{figure}

Three different hyperspectral datasets are used to conduct experiments. The first dataset is \textit{\textbf{Houston 2013}}, which was collected over the University of Houston campus on June, 2012 \cite{debes2014hyperspectral}. The spatial size of it is $349\times 1905$, and the number of spectral bands is 144. Fig.$~$\ref{RGBHS} demonstrates a three-channel image as well as the training and test maps of the Houston 2013 data. As shown in the figure, there exist 15 different classes of land covers. Table$~$\ref{HSData} reports the detailed pixel distributions in each class. The second dataset is \textit{\textbf{Houston 2018}} \cite{Rasti2020feature}, which has a larger spatial size (i.e., $601\times2384$) but less spectral bands (i.e., 48) than the Houston 2013 data. There are 20 different land-cover classes to discriminate.
In this paper, we used the training portion of the whole dataset, which was distributed by the Image Analysis and Data Fusion Technical Committee of the IEEE Geoscience and Remote Sensing Society (GRSS) and the University of Houston for the 2018 data fusion contest \cite{xu2019advanced}.
The detailed number of training as well as test pixels in each class and their spatial distributions are illustrated in Table$~$\ref{HU2018Data} and Fig.$~$\ref{RGBHU2018}, respectively. The last dataset is \textit{\textbf{HyRANK}}$\footnote{http://www2.isprs.org/commissions/comm3/wg4/HyRANK.html}$, which is comprised of two hyperspectral images: \textit{\textbf{Dioni}} and \textit{\textbf{Loukia}}.
The spatial sizes of Dioni and Loukia are $250\times1376$ and $249\times945$, respectively. Both of them contain 176 spectral bands. The available pixels in Dioni are used as the training set, while those in Loukia are used as the test set. Seven common land-cover classes in both images are pre-defined. They are \textit{Dense Urban Fabric}, \textit{Non Irrigated Arable Land}, \textit{Olive Groves}, \textit{Dense Sclerophyllous Vegetation}, \textit{Sparse Sclerophyllous Vegetation}, \textit{Sparsely Vegetated Areas}, and \textit{Water}. Table$~$\ref{HyRANKData} demonstrates the detailed number of available pixels in each class, and Fig.$~$\ref{RGBHyRANK} shows the three-channel images and their pixel distributions.

\begin{table}
  \centering
  \caption{Pixel distributions on the Houston 2013 data. Note that `Percent' means the proportion between the training pixels and the total number of available pixels.}\label{HSData}
  \scalebox{0.9}{
  \begin{tabular}{ccccc}
  \hline
     Class No. & Class Name & Training & Test & Percent\\
     \hline
     \hline
     1 & Healthy grass & 198 & 1053 & 15.83\%\\
     2 & Stressed grass & 190 & 1064 &15.15\%\\
     3 & Synthetic grass & 192 & 505 &27.55\%\\
     4 & Tree & 188 & 1056 &15.11\%\\
     5 & Soil & 186 & 1056 &14.98\%\\
     6 & Water & 182 & 143 &56.00\%\\
     7 & Residential & 196 & 1072 &15.46\%\\
     8 & Commercial & 191 & 1053 &15.35\%\\
     9 & Road & 193 & 1059 &15.42\%\\
     10 & Highway & 191 & 1036 &15.57\%\\
     11 & Railway & 181 & 1054 &14.66\%\\
     12 & Parking lot 1 & 192 & 1041 &15.57\%\\
     13 & Parking lot 2 & 184 & 285 &39.23\%\\
     14 & Tennis court & 181 & 247 &42.29\%\\
     15 & Running track & 187 & 473 &28.33\%\\
     \hline
     \hline
      -  & Total & 2832 & 12197 &18.84\%\\
     \hline
   \end{tabular}
   }
\end{table}

\begin{table}
  \centering
  \caption{Pixel distributions on the Houston 2018 data. Note that `Percent' means the proportion between the training pixels and the total number of available pixels. }\label{HU2018Data}
    \scalebox{0.9}{
    \begin{tabular}{ccccc}
    \hline
    \multicolumn{1}{c}{Class No.} & \multicolumn{1}{c}{Class Name} & \multicolumn{1}{c}{Training} & \multicolumn{1}{c}{Test} & \multicolumn{1}{c}{Percent} \\
    \hline
    \hline
    1     & \multicolumn{1}{c}{Healthy grass} & 1458  & 8341  & 14.88\% \\
    2     & \multicolumn{1}{c}{Stressed grass} & 4316  & 28186 & 13.28\% \\
    3     & \multicolumn{1}{c}{Synthetic grass} & 331   & 353   & 48.39\% \\
    4     & \multicolumn{1}{c}{Evergreen Trees} & 2005  & 11583 & 14.76\% \\
    5     & \multicolumn{1}{c}{Deciduous Trees} & 676   & 4372  & 13.39\% \\
    6     & \multicolumn{1}{c}{Soil} & 1757  & 2759  & 38.91\% \\
    7     & \multicolumn{1}{c}{Water} & 147   & 119   & 55.26\% \\
    8     & \multicolumn{1}{c}{Residential} & 3809  & 35953 & 9.58\% \\
    9     & \multicolumn{1}{c}{Commercial} & 2789  & 220895 & 1.25\% \\
    10    & \multicolumn{1}{c}{Road} & 3188  & 42622 & 6.96\% \\
    11    & \multicolumn{1}{c}{Sidewalk} & 2699  & 31303 & 7.94\% \\
    12    & \multicolumn{1}{c}{Crosswalk} & 225   & 1291  & 14.84\% \\
    13    & \multicolumn{1}{c}{Major Thoroughfares} & 5193  & 41165 & 11.20\% \\
    14    & \multicolumn{1}{c}{Highway} & 700   & 9149  & 7.11\% \\
    15    & \multicolumn{1}{c}{Railway} & 1224  & 5713  & 17.64\% \\
    16    & \multicolumn{1}{c}{Paved Parking Lot } & 1179  & 10296 & 10.27\% \\
    17    & \multicolumn{1}{c}{Gravel Parking Lot } & 127   & 22    & 85.23\% \\
    18    & \multicolumn{1}{c}{Cars} & 848   & 5730  & 12.89\% \\
    19    & \multicolumn{1}{c}{Trains} & 493   & 4872  & 9.19\% \\
    20    & \multicolumn{1}{c}{Seats} & 1313  & 5511  & 19.24\% \\
    \hline
    \hline
     -    & Total & 34477 & 470235 & 6.83\% \\
     \hline
    \end{tabular}%
    }
\end{table}%

\begin{table}
  \centering
  \caption{Pixel distributions on the HyRANK data.}\label{HyRANKData}
  \scalebox{0.9}{
  \begin{tabular}{ccccc}
  \hline
     Class No. & Class Name & Dioni & Loukia \\
     \hline
     \hline
     1 & Dense Urban Fabric & 1262 & 288 \\
     2 &  Non Irrigated Arable Land & 614 & 542 \\
     3 & Olive Groves & 1768 & 1401 \\
     4 & Dense Sclerophyllous Vegetation & 5035 & 3793 \\
     5 & Sparse Sclerophyllous Vegetation & 6374 & 2803 \\
     6 & Sparsely Vegetated Areas & 1754 & 404 \\
     7 & Water & 1612 & 1393 \\
     \hline
     \hline
      -  & Total & 18419 & 10624 \\
     \hline
   \end{tabular}
   }
\end{table}

In order to effectively analyze the performance of the proposed model, we implement two different kinds of experiments on these three data. The first one is to evaluate the effects of different components in the proposed model, including the spectral attention modules and the spatial attention modules. The second one is to compare the proposed model with some state-of-the-art CNN-related models. All the models are simulated by PyTorch on a computer with 32GB RAM and a GTX TITAN X graphic card. The input cube size is empirically set to $11\times11$, because larger sizes will lead to the serious overlap problem \cite{hansch2017correct, lange2018influence}. The optimizer is Adam with default parameters. The learning rate, the batch size, and the training epochs are set to 0.001, 128, and 200, respectively. To alleviate the effects of random initialization on the performance, all the experiments are repeatedly implemented 10 times, and the average performance are recorded. To quantitatively evaluate the performance of each model, we adopt the overall accuracy (OA), the average accuracy (AA), the per-class accuracy, the Kappa coefficient, and the F1 score as indicators.

\subsection{Model Analysis}
\subsubsection{Effects of different components}
\begin{figure*}
  \centering
  \includegraphics[scale=0.55]{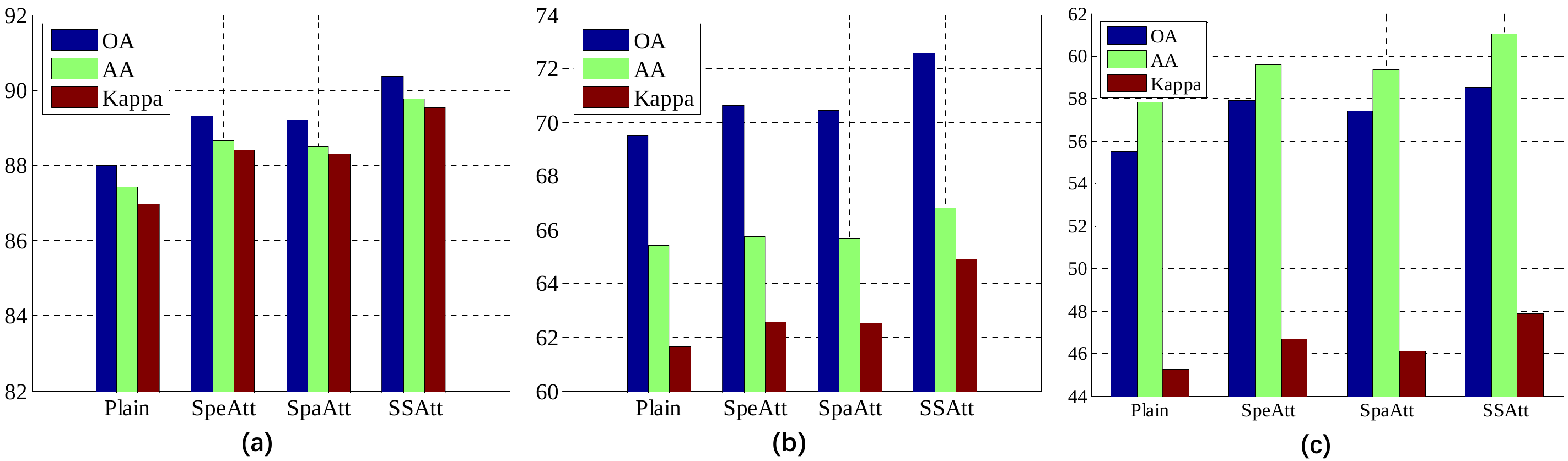}\\
  \caption{Effects of different components on the classification performance. (a) Houston 2013 data. (b) Houston 2018 data. (c) HyRANK data.}\label{EffectsComponents}
\end{figure*}

Different from the traditional CNN models, our proposed model incorporates spectral attention and spatial attention modules into CNN. In this subsection, we test the effectiveness of these attention modules. Specifically, we adopt the convolutional network without any attention modules as a baseline model (abbreviated as \textit{Plain}). It has three convolutional layers and an output layer. The kernel size of each convolutional layer is $3\times3$, and the channel numbers from the first to the third convolution layer are sequentially set to 32, 64, and 128. We compare \textit{Plain} with the spectral attention sub-network (abbreviated
as \textit{SpeAtt}), the spatial attention sub-network (abbreviated as \textit{SpaAtt}), and their integrated two-branch network (abbreviated as \textit{SSAtt}).

Fig.$~$\ref{EffectsComponents} shows the classification performance of different models in terms of OA, AA, and Kappa values. Different colors denote different evaluation indicators. From this figure, we can observe that \textit{SpeAtt} and \textit{SpaAtt} achieve similar performance in most cases, and both of them are better than \textit{Plain} on three data, which certifies the effectiveness of our proposed spectral attention modules and spatial attention modules. In addition, after combining the results of \textit{SpeAtt} and \textit{SpaAtt}, our proposed model \textit{SSAtt} is able to further improve the classification performance. It indicates that \textit{SSAtt} can integrate the complementary information between \textit{SpeAtt} and \textit{SpaAtt}.

\subsubsection{Effects of attention numbers}

\begin{table}
  \centering
  \caption{Effects of different numbers of attention modules on the \textit{SpeAtt} model.}\label{SpeAtt}
  \begin{tabular}{ccccc}
     \hline
     Model & First & Second & Third & OA \\
     \hline
     \hline
     \textit{Plain} & \xmark & \xmark & \xmark & 87.98 \\
     \textit{$SpeAtt_{1}$} & \cmark & \xmark & \xmark  & 88.40 \\
     \textit{$SpeAtt_{1}$} & \xmark & \cmark & \xmark & 88.75 \\
     \textit{$SpeAtt_{1}$} & \xmark & \xmark &\cmark  & 88.53 \\
     \textit{$SpeAtt_{2}$} &\cmark  & \cmark & \xmark & 88.98 \\
     \textit{$SpeAtt_{2}$} & \cmark & \xmark & \cmark & 88.73 \\
     \textit{$SpeAtt_{2}$} & \xmark & \cmark & \cmark & 89.14 \\
     \textit{$SpeAtt_{3}$} & \cmark & \cmark & \cmark & 89.69 \\
     \hline
     \hline
   \end{tabular}
\end{table}

\begin{table}
  \centering
  \caption{Effects of different numbers of attention modules on the \textit{SpaAtt} model.}\label{SpaAtt}
  \begin{tabular}{ccccc}
     \hline
     Model & First & Second & Third & OA \\
     \hline
     \hline
     \textit{Plain} & \xmark & \xmark & \xmark & 87.98 \\
     \textit{$SpaAtt_{1}$} & \cmark & \xmark & \xmark  & 88.34 \\
     \textit{$SpaAtt_{1}$} & \xmark & \cmark & \xmark & 88.77 \\
     \textit{$SpaAtt_{1}$} & \xmark & \xmark &\cmark  & 88.30 \\
     \textit{$SpaAtt_{2}$} &\cmark  & \cmark & \xmark & 89.20 \\
     \textit{$SpaAtt_{2}$} & \cmark & \xmark & \cmark & 88.47 \\
     \textit{$SpaAtt_{2}$} & \xmark & \cmark & \cmark & 89.07 \\
     \textit{$SpaAtt_{3}$} & \cmark & \cmark & \cmark & 89.61 \\
     \hline
     \hline
   \end{tabular}
\end{table}

The last subsection evaluates the effectiveness of different kinds of attention modules, but the number of attention modules may also affect the classification performance. Therefore, in this subsection, we take the Houston 2013 data as an instance, and attempt to comprehensively test the effects of different numbers of attention modules on the classification performance. Since the total number of convolutional layers in the proposed model is three, the maximal number of attention modules is also three. Here, we change the number of attention modules from one to three, and record the classification performance of \textit{SpeAtt} and \textit{SpaAtt} models with different combinations.

Table$~$\ref{SpeAtt} demonstrates the classification performance achieved by \textit{SpeAtt} with different numbers of attention modules. The subscript numbers refer to the number of attention modules used by each model. The symbol `\cmark' represents that the corresponding layer contains an attention module, while the symbol `\xmark' does not. From this table, it can be observed that when the first convolutional layer adopts the spectral attention module, its performance is inferior to both the second and the third convolutional layers. This is because the first convolutional layer has less discriminative features than the other layers. In comparison with the second convolutional layer, the third one contains more discriminative features, but its improvement space is smaller than the second one. Therefore, when equipped with the attention module, the second layer obtains the highest OA. Different from \textit{$SpeAtt_{1}$}, \textit{$SpeAtt_{2}$} adopts attention modules on two layers. Due to the exploitation of more attention modules, \textit{$SpeAtt_{2}$} is able to achieve higher OAs than its \textit{$SpeAtt_{1}$} counterparts. Similarly, \textit{$SpeAtt_{3}$} is better than \textit{$SpeAtt_{2}$}.

Similar to Table$~$\ref{SpeAtt}, Table$~$\ref{SpaAtt} shows the classification performance of \textit{SpaAtt} using different numbers of attention modules. Again, \textit{$SpaAtt_{3}$} obtains the best performance, because every convolutional layer has been refined by the spatial attention modules. In addition, \textit{$SpaAtt_{2}$} is superior to its \textit{$SpaAtt_{1}$} counterparts. For example, \textit{$SpaAtt_{2}$} with attention modules in the first and the second layers is better than the first (i.e., the second row in table$~$\ref{SpaAtt}) and the second \textit{$SpaAtt_{1}$} models. In terms of \textit{$SpaAtt_{1}$} models, the third convolutional layer is inferior to the other ones, because it has less spatial information. Although the first convolutional layer has more spatial information than the second one, its feature representation ability is not good enough. This is why the second \textit{$SpaAtt_{1}$} model achieves higher OA than the first one.

\subsubsection{Analysis on $\alpha$ and $\beta$}
\begin{table}
  \centering
  \caption{Values of $\alpha$ and $\beta$ on different data.}\label{Hyperparameter}
  \scalebox{0.9}{
  \begin{tabular}{ccccc}
    \hline
    Data & Band Number & Spatial Size & $\alpha$ & $\beta$ \\
    \hline
    \hline
    Houston 2013 & 144 & $349\times1905$ & 0.5030 & 0.4970 \\
    Houston 2018 & 48 & $601\times2384$ & 0.4803 & 0.5197 \\
    HyRANK & 176 & $250\times1376$ & 0.5208 & 0.4792 \\
    \hline
    \hline
  \end{tabular}
  }
\end{table}

Since the whole network is optimized by the gradient descent algorithm, $\alpha$ and $\beta$ are also optimized by it. Table$~$\ref{Hyperparameter} shows the final values of $\alpha$ and $\beta$ on three datasets. It is interesting to observe that the optimal $\alpha$ and $\beta$ values vary for different data, because they have different spatial and spectral resolutions. In particular, both Houston 2013 and HyRANK data contain more than 100 spectral bands, which provide more discriminative information than the spatial domain. Therefore, $\alpha$ is relatively larger than $\beta$ for them, especially for the HyRANK data. On the contrary, Houston 2018 only has 48 spectral bands, but its spatial information is rich, making $\beta$ larger than $\alpha$. Based on these observations, adaptively optimizing $\alpha$ and $\beta$ is a better choice than empirically fixing them. Although it takes time to optimize them, empirically choosing them on different data will also cost time.

\begin{table*}
  \centering
  \caption{Classification results (\%) and computation time (seconds) of different models on the Houston 2013 data.}\label{HSResults}
  \scalebox{0.9}{
  \begin{tabular}{ccccccccc}
    \hline
    Class No. & PPF & 2DCNN & ECNN & GCNN & 3DCNN & SSRN & MSDNSA &SSAtt   \\
    \hline
    \hline
    1  & 84.20 & 56.73 & \textbf{87.49} & 87.47  & 78.63 & 81.48 & 82.72 & 82.54 \\
    2  & 96.00 & 64.68 & 80.99 & 86.01 & 93.23 & 92.48 & 99.81 & \textbf{99.92}	\\
    3 & \textbf{98.61} & 44.67 & 87.72 & 78.22 & 40.99 & 98.02 & 89.70 & 86.48 \\
    4 & 94.89 & 59.05 & 90.43 & 85.02 & 97.44 & 98.11 & 95.08 & \textbf{99.57}  \\
    5 & 96.67 & 68.31 & \textbf{100} & 99.89 & 87.31 & 99.91 & 94.89 & 99.61 \\
    6 & 82.94 & 70.21 & \textbf{97.90} & 89.44 & 79.02 & 95.80 & 95.80 & 83.71 \\
    7 & 82.67 & 82.57 & 90.48 & 90.19 & \textbf{90.49} & 89.46 & 85.63 & 89.92 \\
    8 & 52.69 & 52.12 & 58.51 & 74.44 & 59.83 & 69.90 & \textbf{85.57} & 81.94  \\
    9 & 78.21 & 70.35 & 79.77 & 84.42 & 81.11 & 84.04 & \textbf{86.02} & 85.99 \\
    10 & 72.78 & 60.37 & 64.28 & 63.61 & 69.59 & 82.34 & 60.33 & \textbf{88.81}  \\
    11 & 87.38 & 72.16 & 78.37 & 80.06 & 75.14 & \textbf{93.17} & 87.67 & 90.53 \\
    12 & 80.52 & 44.63 & 78.29 & 87.30 & 82.23 & \textbf{90.80} & 90.78 &  89.48 \\
    13 & 70.88 & 87.02 & 76.84 & 85.06 & 82.11 & 72.98 & \textbf{90.88} & 86.81 \\
    14 & 99.51 & 96.92 & 99.19 & \textbf{100} & 80.57 & 99.19 & 99.60 & 95.34  \\
    15 & \textbf{94.71} & 14.93 & 77.04 & 56.95 & 39.11 & 94.08 & \textbf{94.71} & 85.77 \\
    \hline
    \hline
    OA  & 83.84 & 61.85 & 84.04 & 84.12 & 78.19 & 89.46 & 87.78 & \textbf{90.38} \\
    AA & 84.84 & 62.98 & 83.33 & 82.94 & 75.79 & 89.45 & 89.28 & \textbf{89.76} \\
    Kappa & 82.53 & 58.64 & 82.54 & 82.51 & 76.27 & 88.58 & 86.73 & \textbf{89.55} \\
    F1 &85.72&60.92&82.35&81.27&76.79&88.03&88.69&\textbf{90.67} \\
    Training(s) & 779.47 & \textbf{21.46} & 21.55 & 22.54 & 11472.14 & 715.44 & 4994.82 & 180.36 \\
    Test(s) & 0.30 & \textbf{0.12} & 0.13 & 0.16 & 77.21 & 5.10 & 120.12 & 0.78 \\
    \hline
  \end{tabular}
  }
\end{table*}

\begin{figure*}
  \centering
  \includegraphics[scale = 0.6]{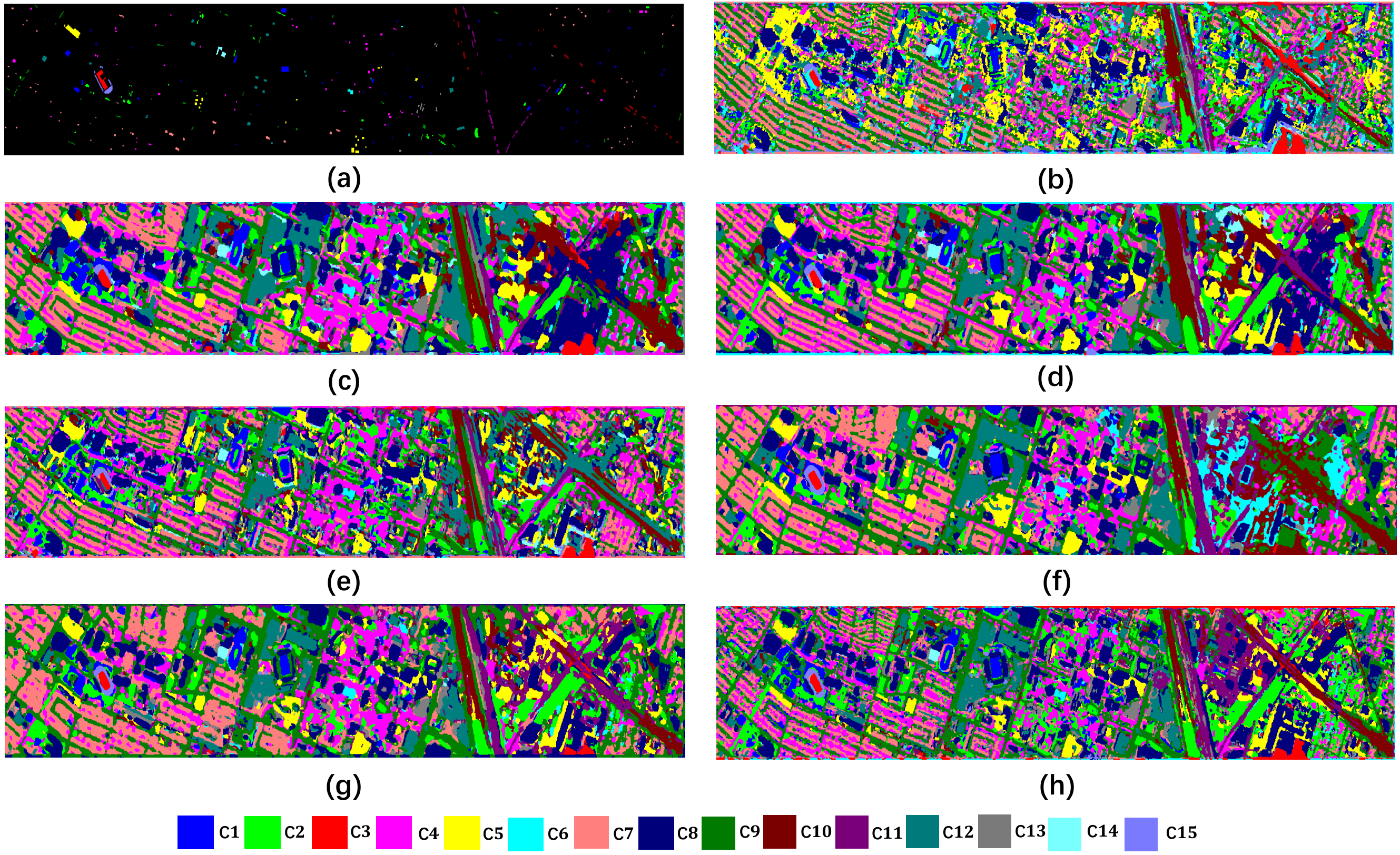}\\
  \caption{Classification maps achieved by seven different models on the Houston 2013 data. (a) Test data map. (b) 2DCNN. (c) ECNN. (d) GCNN. (e) 3DCNN. (f) SSRN. (g) MSDNSA. (h) SSAtt.}\label{HSMaps}
\end{figure*}

\begin{figure*}
  \centering
  \includegraphics[scale = 0.6]{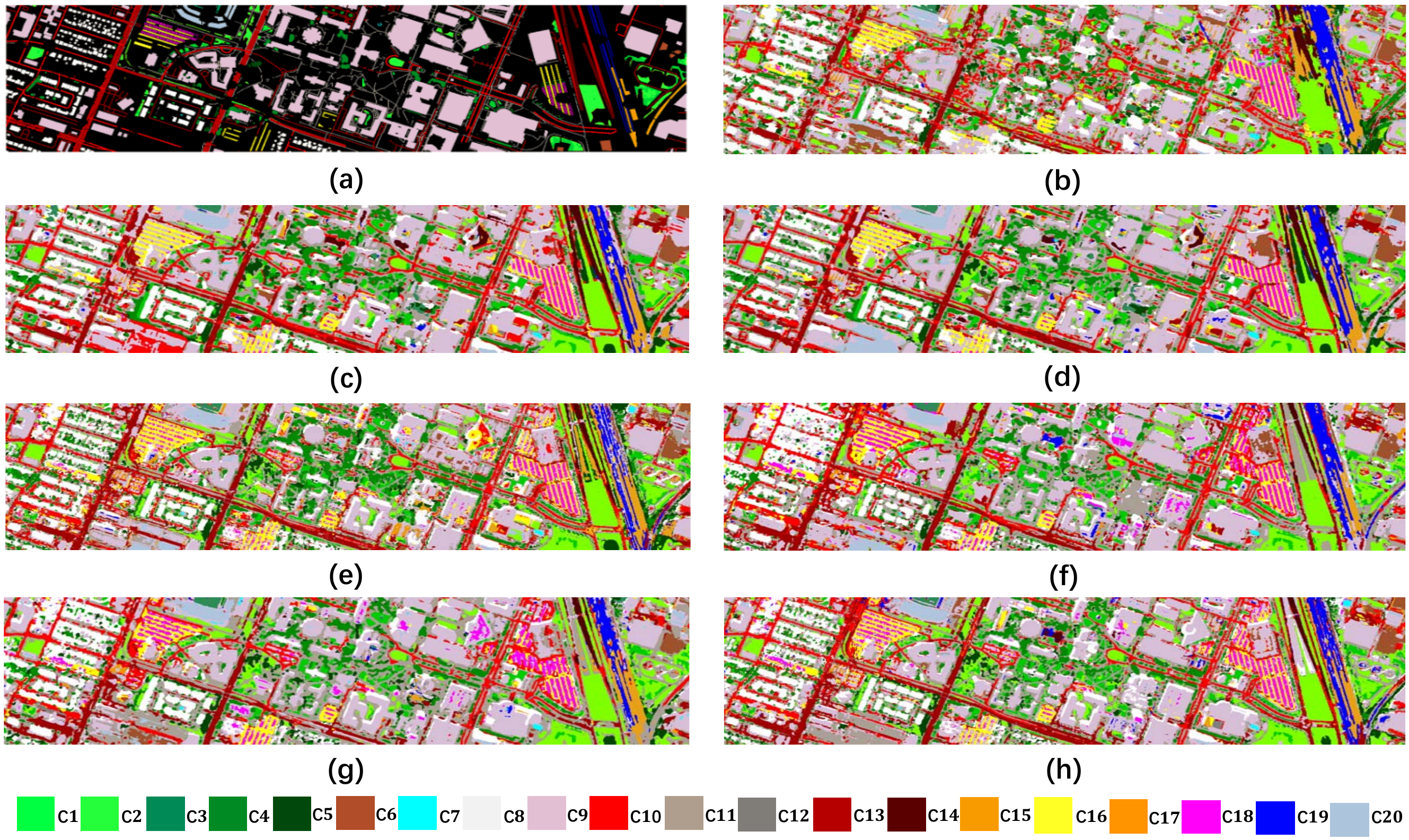}\\
  \caption{Classification maps achieved by seven different models on the Houston 2018 data. (a) Test data map. (b) 2DCNN. (c) ECNN. (d) GCNN. (e) 3DCNN. (f) SSRN. (g) MSDNSA. (h) SSAtt.}\label{HS2018Maps}
\end{figure*}

\begin{table*}
  \centering
  \caption{Classification results (\%) and computation time (seconds) of different models on the Houston 2018 data.}\label{HU2018Results}
  \scalebox{0.9}{
  \begin{tabular}{ccccccccc}
    \hline
    Class No. & 2DCNN & ECNN & GCNN & 3DCNN & SSRN & MSDNSA &SSAtt   \\
    \hline
    \hline
    1  & 20.72 &66.62&54.71 & 58.76 & \textbf{71.38} & 63.28  &  68.83  \\
    2  & 62.45 &75.60 &85.06 & 90.06 & 81.88 & 88.83  & \textbf{90.54} 	\\
    3 & 44.48  &89.52&86.40 & 92.35 & 98.87 & \textbf{99.15} &  88.10 \\
    4 & 83.60  &95.46 &94.16 & \textbf{97.78} & 91.70 & 95.53  & 95.49  \\
    5 & 42.18  &41.72 &50.09 & \textbf{55.95} & 38.76 & 49.89 &  55.10 \\
    6 & 28.63  &32.58 &33.06 & 31.10 & \textbf{40.40} & 38.35  &  30.12 \\
    7 & 0     &0 &\textbf{31.09} & 29.41 & 24.16 & 30.25  & 30.25 \\
    8 & 87.93 &86.44 &82.22 & \textbf{89.30} & 82.74 & 83.56  &  86.37  \\
    9 & 53.44 &64.10 &65.67 & 58.74 & 75.76 & 73.74  & \textbf{77.39} \\
    10 & 41.77 &\textbf{61.46} &57.09 & 55.41 & 54.86 & 45.40  & 54.74   \\
    11 & 45.35 &59.02 &66.04 & 60.55 & 65.46 & \textbf{67.46}  &  63.32 \\
    12 & 2.87 &1.47 &7.67 & \textbf{14.41} & 9.39 & 11.46 & 12.70  \\
    13 & 44.25 &50.36 &51.13 & \textbf{55.18} & 50.76 & 47.96 & 50.34  \\
    14 & 41.84 &\textbf{84.06} &74.49 & 57.55 & 34.43 & 35.38  & 46.91   \\
    15 & 60.21 &\textbf{69.19}&68.28 & 67.01 & 67.73 & 64.76  & 65.31  \\
    16 & 79.98 &88.33 &\textbf{89.28} & 87.10 & 74.18 & 77.82  & 85.01 \\
    17 & \textbf{100} &95.45 &\textbf{100} & \textbf{100} & \textbf{100} &  \textbf{100} &  \textbf{100} \\
    18 & 44.10 &38.03 &43.23 & 57.03 & \textbf{79.55} & 73.30  & 66.58 \\
    19 & \textbf{95.87} &90.68 &92.34 & 85.76 & 85.09 & 80.17 & 94.68 \\
    20 & 41.23 &70.57 &\textbf{85.96} & 67.50 & 61.67 & 81.98  & 74.14 \\
    \hline
    \hline
    OA  & 54.59 &65.99 &67.05 & 64.19 & 70.52 & 69.30  & \textbf{72.57}  \\
    AA & 51.04 &63.03 &65.89 & 65.55 & 64.44 & 65.41  &  \textbf{66.80} \\
    Kappa & 44.97 &57.64 &59.04 & 56.24 & 62.39 & 61.01  & \textbf{64.89} \\
    F1 & 39.81  &50.15 &50.67  &51.38 &54.69  &54.49  &\textbf{58.02} \\
    Training(s) & \textbf{242.76} &265.33&280.56  & 6193.41 & 1155.94 & 23685.47 & 1060.20  \\
    Test(s) & \textbf{4.66} &5.20&6.12  & 132.18 & 51.49 & 1526.66 & 16.77  \\
    \hline
  \end{tabular}
  }
\end{table*}

\begin{table*}
  \centering
  \caption{Classification results (\%) and computation time (seconds) of different models on the HyRANK data.}\label{HyRANKResults}
  \scalebox{0.9}{
  \begin{tabular}{ccccccccc}
    \hline
    Class No. & 2DCNN & ECNN & GCNN & 3DCNN & SSRN & MSDNSA &SSAtt   \\
    \hline
    \hline
    1  &44.79  &17.01  &35.07   &45.14  &3.82  &59.03  &\textbf{82.99}  \\
    2  &\textbf{75.28}  &0.92  &39.30   &74.54  &9.41  &54.43  &21.59\\
    3  &\textbf{47.39}  &0.21  &12.21   &36.12  &10.71  &0  &30.34\\
    4   &66.86  &\textbf{94.99}  &88.53   &73.64  &76.43  &68.05  &60.93\\
    5   &3.28  &2.64  &4.07   &3.14  &42.99  &\textbf{51.02}  &50.16\\
    6  &59.65  &\textbf{86.14}  &61.63   &50.99  &70.30  &4.70  &81.44\\
    7   &\textbf{100}  &\textbf{100}  &\textbf{100}   &\textbf{100}  &\textbf{100}  &\textbf{100}  &\textbf{100}\\
    \hline
    \hline
    OA  &51.42  &51.53  &52.70   &51.96  &56.41  &55.42  & \textbf{58.55}\\
    AA   &56.75  &43.13  &48.69   &54.79  &44.81  &48.17  &\textbf{61.06}\\
    Kappa  &40.78  &34.60  &37.96   &40.91  &43.97  &43.54  &\textbf{47.88}\\
    F1 &45.49  &31.78  &40.60  & 44.99 & 40.12 &40.61  & \textbf{51.61}  \\
    Training(s)  &\textbf{136.61}&140.51&141.05&17326.13 &8474.97 &37695.25 &272.56\\
    Test(s)   &\textbf{0.059}  &0.076  &0.080   &97.62 & 8.58 &122.94  &0.23 \\
    \hline
  \end{tabular}
  }
\end{table*}

\begin{figure*}
  \centering
  \includegraphics[scale = 0.6]{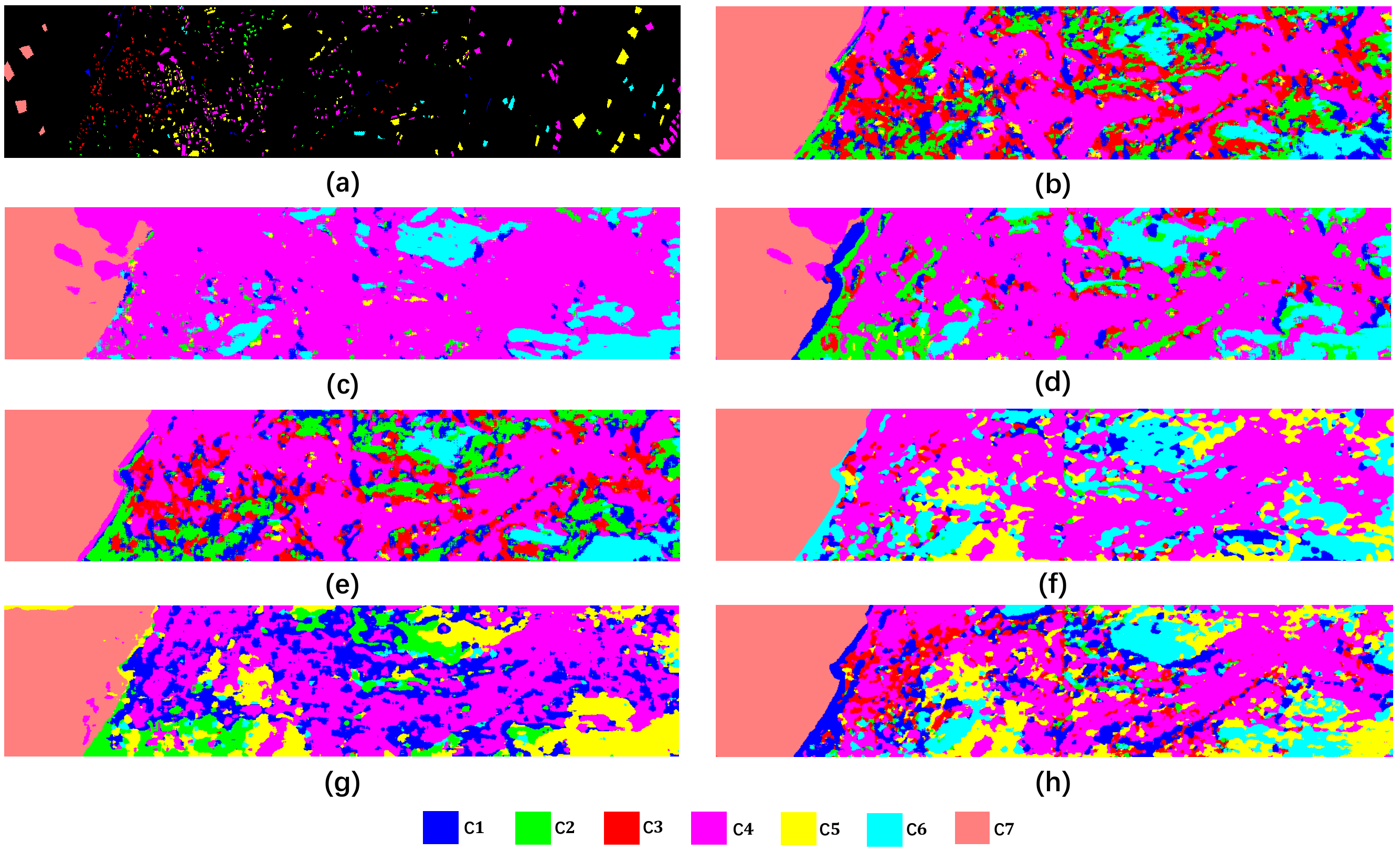}\\
  \caption{Classification maps achieved by seven different models on the HyRANK data. (a) Test data map. (b) 2DCNN. (c) ECNN. (d) GCNN. (e) 3DCNN. (f) SSRN. (g) MSDNSA. (h) SSAtt.}\label{HyRANKMaps}
\end{figure*}

\subsection{Model Comparison}
Since our proposed model \textit{SSAtt} is based on CNN, we compare it with seven state-of-the-art CNN-related models to evaluate its performance. These models include Pixel-Pair CNN (\textit{PPF}) in \cite{li2017hyperspectral}, \textit{2DCNN} in \cite{chen2016deep}, Attribute Profile based CNN (\textit{ECNN}) in \cite{aptoula2016deep}, Gabor Filtering based CNN (\textit{GCNN}) in \cite{chen2017hyperspectral}, \textit{3DCNN} in \cite{chen2016deep}, Spectral-Spatial Residual Network (\textit{SSRN}) in \cite{zhong2018spectral}, and Dense Convolutional Networks with Spectral-Wise Attention Mechanism (\textit{MSDNSA}) in \cite{fang2019hyperspectral}. For these models, we transfer the network structures from the original papers and re-implement them by ourselves on the three datasets.

\subsubsection{Quantitative comparisons}
Table$~$\ref{HSResults} demonstrates the classification performance in terms of OA, AA, Kappa, F1 score as well as each class accuracy achieved by different models on the Houston 2013 data. Note that the numbers reported in bold type face indicate the best results in each row. From this table, several conclusions can be derived. First of all, \textit{PPF} is able to achieve satisfactory classification results because of its deep network structure (i.e., 10 convolutional layers) along with the novel data augmentation strategy. However, \textit{PPF} is a spectral classification model, which ignores the use of spatial information, making it difficult to discriminate land-cover classes with similar materials. For instance, the `\emph{Commercial}' class whose accuracy is only 52.69\% can be easily misclassified as `\emph{Residential}'. Second, for the 2-D convolutional networks (i.e., \textit{2DCNN}, \textit{ECNN}, and \textit{GCNN}), \textit{2DCNN} obtains inferior performance in most classes when compared to \textit{ECNN} and \textit{GCNN}. This can be explained by the loss of discriminative spectral information in the \textit{2DCNN} model. Instead of using the first principal component as input in \textit{2DCNN}, \textit{ECNN} and \textit{GCNN} adopt more principal components as inputs and extract some spatial features from them, thus improving the classification performance. Third, for the 3-D convolutional networks (i.e., \textit{3DCNN}, \textit{SSRN}, and \textit{MSDNSA}), \textit{3DCNN} significantly improves the performance of \textit{2DCNN}, but the amount of improvement is still not as good as what we expected. One of the possible reasons is that \textit{3DCNN} has a large number of learned parameters, while the available training samples are not large enough to train it. Another possible reason is that the convolutional features are not fully exploited. In comparison with \textit{3DCNN}, \textit{SSRN} adopts smaller convolutional kernels to reduce the number of training parameters and residual structures to combine features from different convolutional layers; \textit{MSDNSA} uses densely-connected structures and spectral attention modules to integrate and refine convolutional features, respectively. Therefore, they are capable of improving the performance of \textit{3DCNN} by a large margin in terms of OA, AA, Kappa, and F1 score. The last but not the least, for the seven comparison models, \textit{SSRN} is able to achieve the best OA, AA, and Kappa values. Benefiting from the designed attention modules, which enhance the discriminative information and suppress the unnecessary information in spectral and spatial domains, our proposed model \textit{SSAtt} can further improve these values, which certifies the effectiveness of it.

Table$~$\ref{HU2018Results} and Table$~$\ref{HyRANKResults} compare the classification performance of different models on the Houston 2018 data and the HyRANK data, respectively. Note that we do not re-implement the \textit{PPF} model on these two datasets, because the pixel-pair method will generate more than 10 millions of training samples, which is out of the computation capability of our GPU. The same as Table$~$\ref{HSResults}, some similar conclusions can be observed from Table$~$\ref{HU2018Results} and Table$~$\ref{HyRANKResults}. For 2-D convolutional networks, \textit{2DCNN} does not work as well as the other two models in terms of OA. For the 3-D convolutional networks, \textit{3DCNN} achieves inferior classification results in terms of OA and Kappa values as compared to \textit{SSRN} and \textit{MSDNSA}. When comparing 2-D convolutional networks and 3-D convolutional networks, it can be found that \textit{SSRN} is a relatively better model in terms of  OA and Kappa values. In comparison with \textit{SSRN}, the \textit{SSAtt} model is capable of improving the OA, AA, Kappa, and F1 scores on both datasets. These conclusions can sufficiently validate the effectiveness of the proposed model.

\subsubsection{Qualitative comparisons}
In addition to the quantitative results in Table$~$\ref{HSResults}-Table$~$\ref{HyRANKResults}, we also demonstrate the classification maps acquired by seven different models qualitatively. Fig.$~$\ref{HSMaps}, Fig.$~$\ref{HS2018Maps}, and Fig.$~$\ref{HyRANKMaps} represent the classification maps on the Houston 2013 data, the Houston 2018 data, and the HyRANK data, respectively. In these figures, different colors correspond to different land-cover classes. When comparing the classification maps in sub-figures (b)-(h) with the ground-truth map in sub-figure (a), it can be observed that the proposed model \textit{SSAtt} obtains more reasonable maps than the other comparison models, which indicates its superiority. However, all of the classification maps in sub-figures (b)-(h) seem to be a little over-smoothed. This is caused by the fact that the inputs of these CNN-related models are cubes around each pixels, making the boundary pixels between two objects easily misclassified.

\subsubsection{Computation time}

In Table$~$\ref{HSResults}-Table$~$\ref{HyRANKResults}, the last two rows record the training time and test time of different models. Without loss of generality, it can be observed that the training stage costs much more time than the test stage for each model. Specifically, \textit{2DCNN} is the most efficient model, because it only processes one component extracted from the whole spectral bands. Nevertheless, its classification performance is significantly inferior to the 3-D CNN-related models (i.e., \textit{3DCNN}, \textit{SSRN}, and \textit{MSDNSA}) and the proposed \textit{SSAtt} model in most cases. In contrary, although the 3-D CNN-related models are generally able to achieve satisfactory results, their computation costs are very high due to the simultaneous convolution operators in both spectral and spatial domains. Take Houston 2013 data as an example, it takes more than ten thousand seconds to train the \textit{3DCNN} model, and about five thousand seconds to train the \textit{MSDNSA} model, while other models only costs hundreds of seconds to train. Different from 3-D CNN-related models, \textit{PPF} only deals with the spectral information, but it still takes much time to train on the Houston 2013 data. This is caused by its pair-wise classification strategy which increases the number of available training samples exponentially. In summary, \textit{SSRN} has the best balance between the computation time and the classification performance among the seven compared models. In comparison with the \textit{SSRN} model, our proposed model \textit{SSAtt} takes less time to train and test on all of the three datasets.

\section{Conclusions}
In this paper, we proposed a hyperspectral image classification method using an attention aided CNN model. We firstly designed two different classes of attention modules (i.e., the spectral attention module and the spatial attention module) using some convolutional layers. Then, we incorporated them into the original CNN to construct a spectral attention sub-network and a spatial attention sub-network, which can focus on more discriminative information in the spectral domain and the spatial domain, respectively. Finally, an integrated method was employed to fuse the complementary information from these two sub-networks.
In order to validate the effectiveness of the proposed model, we constructed two kinds of experiments on three different data. The first one analyzed the effects of different attention modules. The other one compared the performance of the proposed model with several state-of-the-art CNN-related models. The experimental results show that both the spectral attention sub-network and the spatial attention sub-network are able to obtain higher performance than the original CNN with the aid of attention modules, and their integrated model can further improve the performance. In comparison with the state-of-the-art models, the proposed model is able to achieve the best performance in terms of OA, AA, Kappa, and F1 scores, and has a good balance between the classification performance and computation time.

\section{Acknowledgments}
We would like to thank the ISPRS for providing the HyRANK data, and the National Center for Airborne Laser Mapping, the Hyperspectral Image Analysis Laboratory at the University of Houston, and the IEEE GRSS Image Analysis and Data Fusion Technical Committee for providing the Houston data sets.

\bibliography{IEEEfull,SSAttention}
\bibliographystyle{IEEEbib}

\end{document}